\documentclass[prb,showpacs,superscriptaddress,twocolumn,preprintnumbers,amsmath,amssymb]{revtex4-1}

\usepackage[dvipdfmx]{graphicx}
\usepackage{dcolumn}
\usepackage{bm}
\usepackage{amsmath}
 
\graphicspath{{graphics/}}  
\usepackage{color}
\usepackage{cancel}
\usepackage{ulem}

\begin{document}


\title{On orbital angular momentum conservation \\ in Brillouin light scattering within a ferromagnetic sphere}

\author{A.~Osada}
\email{alto@iis.u-tokyo.ac.jp}
\email{Current affiliation: Institute for Nano Quantum Information Electronics, The University of Tokyo, Meguro-ku, Tokyo 153-8505, Japan.}
\author{A.~Gloppe}
\affiliation{Research Center for Advanced Science and Technology (RCAST), The University of Tokyo, Meguro-ku, Tokyo 153-8904, Japan}
\author{Y.~Nakamura}
\affiliation{Research Center for Advanced Science and Technology (RCAST), The University of Tokyo, Meguro-ku, Tokyo 153-8904, Japan}
\affiliation{Center for Emergent Matter Science (CEMS), RIKEN, Wako, Saitama 351-0198, Japan}
\author{K.~Usami}
\email{usami@qc.rcast.u-tokyo.ac.jp}
\affiliation{Research Center for Advanced Science and Technology (RCAST), The University of Tokyo, Meguro-ku, Tokyo 153-8904, Japan}
\date{\today}

\begin{abstract}
Magnetostatic modes supported by a ferromagnetic sphere have been known as the Walker modes, each of which possesses an orbital angular momentum as well as a spin angular momentum along a static magnetic field. The Walker modes with non-zero orbital angular momenta exhibit topologically non-trivial spin textures, which we call \textit{magnetic quasi-vortices}. Photons in optical whispering gallery modes supported by a dielectric sphere possess orbital and spin angular momenta forming \textit{optical vortices}.  Within a ferromagnetic, as well as dielectric, sphere, two forms of vortices interact in the process of Brillouin light scattering.  We argue that in the scattering there is a selection rule that dictates the exchange of orbital angular momenta between the vortices. The selection rule is shown to be responsible for the experimentally observed nonreciprocal Brillouin light scattering. 

\begin{description}
\item[PACS numbers]
\end{description}
\end{abstract}

\maketitle


\section{Introduction}
The coupling between electron spins in solids and light is in general very weak. This is because the coupling is inevitably mediated by the orbital degree of the electrons and is realized through spin-orbit interaction for orbits and spins and electric-dipole interaction for orbits and light, respectively~\cite{SB1966}. Although it is possible to coherently (non-thermally) manipulate collective excitations of spins in spin-ordered materials by means of ultrafast optics, where the electric field density of an optical pulse is high both temporally and spatially~\cite{KKR2010,Kimel2005,Satoh2012,Ogawa2015}, an attempt to realize coherent optical manipulation of magnons \textit{in the quantum regime} is hindered by the weakness of the spin-light coupling~\cite{Hisatomi2016}. Given the encouraging development of \textit{circuit} quantum magnonics, where \textit{microwave} photons and magnons are strongly coupled, enabling a coherent energy exchange at the single-quantum level~\cite{Tabuchi2014,Tabuchi2015,Quirion2017}, the similar energy exchange between \textit{optical} photons and magnons has been anticipated. 

To overcome the weakness of the spin-light interaction, \textit{cavity optomagnonics} has been investigated~\cite{Haigh2015, Osada2016, Tang2016, Haigh2016, Kusminskiy2016, Liu2016, SBB2017}. In cavity optomagnonics, the density of states of optical modes are engineered with an optical cavity to enhance spin-light interaction. In particular, spheres of ferromagnetic insulators supporting whispering gallery modes (WGMs) for photons and a spatially uniform magnetostatic mode, called the Kittel mode, for magnons are used as a platform of the cavity optomagnonics. With spheres made of typical ferromagnetic insulator, yttrium iron garnet (YIG), the pronounced sideband asymmetry~\cite{Osada2016, Tang2016, Haigh2016}, the nonreciprocity~\cite{Osada2016}, and the resonant enhancement~\cite{Tang2016, Haigh2016} of magnon-induced Brillouin scattering have been demonstrated.

In this context, it is interesting to examine the behavior of magnetostatic modes beyond the simplest Kittel mode. The magnetostatic modes residing in a ferromagnetic sphere under a uniform static magnetic field are known as the Walker modes~\cite{Walker1957, FB1959}. They exhibit, in general, topologically non-trivial spin textures about the axis along the applied magnetic field and might be called \textit{magnetic quasi-vortices}. The magnetic quasi-vortices can be characterized by their \textit{orbital} angular momenta along the symmetry axis~\cite{Bauer2013,Tchernyshyov2015}. Photons in optical whispering gallery modes possess not only spin angular momenta but also orbital angular momenta, too, which echoes the concept known as \textit{optical vortices}~\cite{Allen}. Within the ferromagnetic sphere, the optical vortices can interact with the magnetic quasi-vortices in the course of the Brillouin light scattering. The total orbital angular momentum is then expected to be conserved as long as the symmetry axis of the WGMs coincides with that of the Walker modes, imposing a selection rule on the Brillouin scattering processes. 

In this article, the Brillouin scattering hosted in a ferromagnetic sphere is theoretically investigated putting a special emphasis on the orbital angular momentum exchange between the optical vortices and the magnetic quasi-vortices.  We establish a selection rule imposed by the orbital angular momentum conservation for the Brillouin scattering hosted in a ferromagnetic sphere. The experimentally observed Brillouin scattering by various Walker modes reported in the accompanying paper~\cite{PRL}, which reveals that the scattering is either nonreciprocal or reciprocal depending on the orbital angular momentum of the magnetic quasi-vortices, is then analyzed with the theory developed here and found to be explained well. The result would provide a new area for chiral quantum optics~\cite{Lodahl2017} and topological photonics~\cite{Taylor2014,Soljacic2014} based on optical vortices and magnetic quasi-vortices.

\section{Orbital angular momenta}

\begin{figure}[h]
\includegraphics[width=8.6cm]{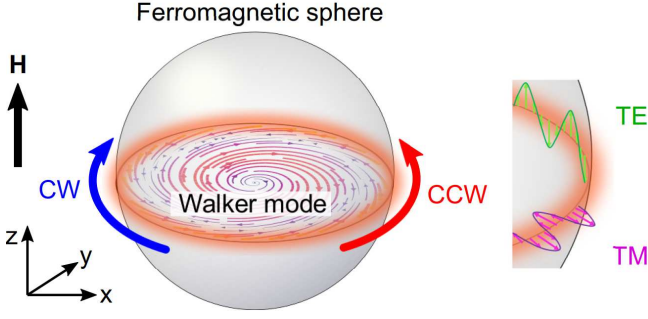}
\caption{\label{Fig:scheme} Schematics of the cavity optomagnonic system considered in this article. A clockwise (CW) or a counterclockwise (CCW), transverse electric (TE) whispering gallery mode (WGM) or transverse magnetic (TM) WGM (see right inset) is Brillouin-scattered by the Walker mode depicted by the field lines.  Here the distribution of the transverse magnetization of the $(4, 0, 1)$ Walker mode on the equatorial plane is shown as an example. The Walker modes and the WGMs are assumed to share the symmetry axis ($z$-axis) along a static magnetic field $H$.}
\end{figure}

The schematics of the cavity optomagnonic system we investigate is shown in Fig.~\ref{Fig:scheme}, where the Walker mode and the WGMs share the symmetry axis ($z$-axis) along a static magnetic field $H$. The Walker modes and the WGMs generally exhibit nonzero orbital angular momenta.  In this section we analyze the orbital angular momenta of these modes.

\subsection{Orbital angular momenta of Walker modes}

\begin{figure}[b]
\includegraphics[width=8.6cm]{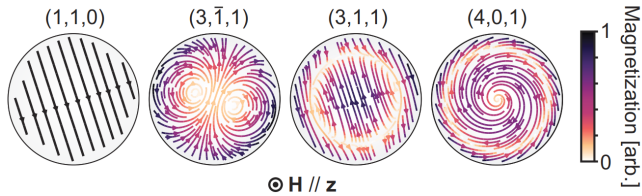}
\caption{\label{Fig4} Transverse magnetization distributions on the equatorial plane of $(1, 1, 0)$, $(3, \bar{1}, 1)$, $(3, 1, 1)$ and $(4, 0, 1)$ Walker modes, whose orbital angular momenta are $\mathcal{L}_{z}^{(1)}=0$, $\mathcal{L}_{z}^{(-1)}=2$, $\mathcal{L}_{z}^{(1)}=0$, and $\mathcal{L}_{z}^{(0)}=1$, corresponding to the winding numbers of the respective spin textures of the Walker modes. Note that the magnetic field is applied parallel to $z$ axis.}
\end{figure}

The orbital angular momentum density $l_{z}^{(m_{\mathrm{mag}})}$ of a magnon along the static magnetic field $\mathbf{H}$ ($\| z$-axis) can be deduced from the dependence of the transverse magnetizations, $M_{x}(t)$ and $M_{y}(t)$, on the azimuthal angle $\phi$ as~\cite{Bauer2013,Tchernyshyov2015}
\begin{equation}
M_{\pm}(t) = M_{x}(t) \pm iM_{y}(t) = M_{\perp} (t) e^{\mp i l_{z}^{(m_{\mathrm{mag}})} \phi}, \label{eq:M}
\end{equation}
where $M_{\perp}(t)=\sqrt{M_{x}^{2}(t)+M_{y}^{2}(t)}$. As for the Walker mode with the index $(n,m_{\mathrm{mag}},r)$~\cite{Walker1957,FB1959} the orbital angular momentum $\mathcal{L}_{z}^{(m_{\mathrm{mag}})}$ can be given by the volume integral of $l_{z}^{(m_{\mathrm{mag}})}$ over the entire sphere and depends on the index $m_{\mathrm{mag}}$, that is, 
\begin{eqnarray}
\mathcal{L}_{z}^{(m_{\mathrm{mag}})} &=& \int  l_{z}^{(m_{\mathrm{mag}})} dV \nonumber \\
&\approx& -\left( m_{\mathrm{mag}}-1\right). \label{eq:Lz_m}
\end{eqnarray}

While the Kittel mode [$(1,1,0)$ mode] has no orbital angular momentum, $\mathcal{L}_{z}^{(1)}=0$, $(4,0,1)$ and $(3,\bar{1},1)$ modes, for instance, have $\mathcal{L}_{z}^{(0)} \approx 1$ and $\mathcal{L}_{z}^{(-1)} \approx 2$, respectively. The approximation in the last line of Eq.~(\ref{eq:Lz_m}) is due to the dipolar interaction with broken axial symmetry.  As the applied static magnetic field $\mathbf{H}$ approaches infinity, the Zeeman energy becomes dominant over the dipole interaction energy, and thus ``$\approx$" becomes ``$=$" in Eq.~(\ref{eq:Lz_m}). Note also that for the Walker modes with $n=m_{\mathrm{mag}}$ and $n=m_{\mathrm{mag}}+1$, Eq.~(\ref{eq:Lz_m}) is exact. We call the Walker modes with non-zero $\mathcal{L}_{z}^{(m_{\mathrm{mag}})}$ as \textit{magnetic quasi-vortices}. The prefix ``\textit{quasi-}" emphasizes the fact that the orbital angular momentum we defined in Eq.~(\ref{eq:Lz_m}) is the approximated one and the fact that magnons are quasi-particle with finite lifetime.

Figure~\ref{Fig4} shows the spatial distributions of the transverse magnetizations for the representative Walker modes $(1, 1, 0)$, $(3, \bar{1}, 1)$, $(3, 1, 1)$, and $(4, 0, 1)$. The modes having non-zero $\mathcal{L}_{z}$ [e.g., $(3, \bar{1}, 1)$ and $(4, 0, 1)$ in Fig.~\ref{Fig4}] exhibit the topologically non-trivial spin textures. Note that the orbital angular momentum $\mathcal{L}_{z}$ here plays a similar role as the \textit{winding number} or the \textit{skyrmion number} in other literature~\cite{KN2016}. 

\subsection{Orbital angular momentum of WGMs}

\begin{figure}[t]
\includegraphics[width=8.6cm]{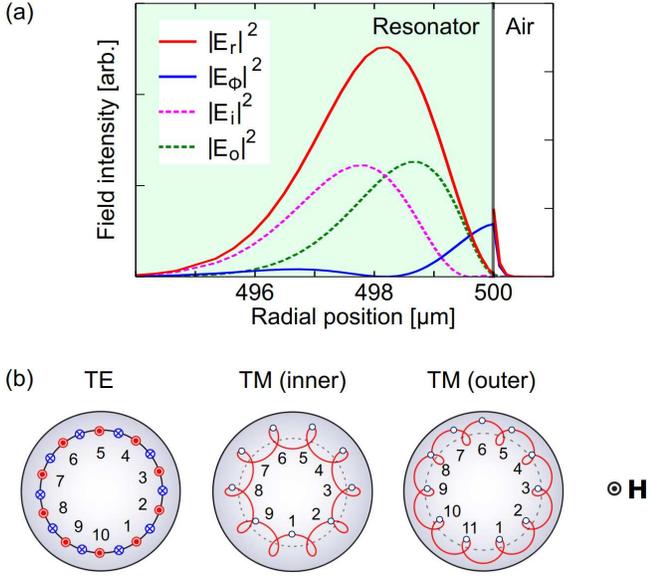}
\caption{\label{Fig:WGM} (a) Intensity profiles of the radial and azimuthal components on the equatorial plane, $\left|E_{r}\right|^{2}$ and $\left|E_{\mathrm{\phi}}\right|^{2}$, of the electric field of the fundamental TM WGM are shown in the red and blue solid lines, respectively. It is assumed that the sphere diameter is $1$~mm and the refractive index is $2.19$ as for YIG. The radial distributions of the \textit{inner} and \textit{outer} components, $\left|E_{i}\right|^{2}$ and $\left|E_{o}\right|^{2}$, are shown in the green and magenta dotted lines. The inner component $E_{i}$ is associated with $\sigma_+$ ($\sigma_-$) component, while the outer component $E_{o}$ is associated with $\sigma_-$ ($\sigma_+$) component for the CCW (CW) orbit. The vertical line indicates the resonator-air boundary. (b) Schematic representation of the difference of orbital angular momenta for the TE, inner TM and outer TM component of a WGM with $m = 10$. The trajectories of the head of the polarization vector for each electric field are shown. The hollow circles around which the electric field directs upward are for the ease of counting the numbers of the rotations of the head. When the mode index is $10$, the orbital angular momentum reads 10, 9 and 11 for the TE, inner TM, and outer TM components, respectively.}
\end{figure}

The electric field of the WGM in an axially symmetric dielectric material has been extensively studied~\cite{BohrenHuffman}. Now, for simplicity, we focus on the azimuthal mode index $m$ which characterizes the azimuthal profile of the electric field of the fundamental WGM. In the spherical basis $\left\{ \Hat{\bm{e}}_{+}, \Hat{\bm{e}}_{0}, \Hat{\bm{e}}_{-} \right\} = \left\{ -\frac{1}{\sqrt{2}} \left( \Hat{\bm{e}}_{x} + i \Hat{\bm{e}}_{y} \right), \Hat{\bm{e}}_{z},\frac{1}{\sqrt{2}} \left( \Hat{\bm{e}}_{x} - i \Hat{\bm{e}}_{y} \right) \right\}$, the electric field of the WGMs of the counterclockwise (CCW) orbit can be written as
\begin{eqnarray}
\bm{E}^{(\mathrm{TE})} \!\! &=& \left( E^{(m)} e^{-im\phi} \right) \Hat{\bm{e}}_{0}^{*}  \label{eq:TE} \\
\bm{E}^{(\mathrm{TM})} \!\! &=& \left( E_{i}^{(m)} e^{-im\phi} \right) e^{i \phi} \Hat{\bm{e}}_{+}^{*} \notag \\ &&\hspace{10mm}- \left( E_{o}^{(m)} e^{-im\phi} \right) e^{-i \phi} \Hat{\bm{e}}_{-}^{*}. \label{eq:TM}
\end{eqnarray}
where $\bm{E}^{(\mathrm{TE})}$ and $\bm{E}^{(\mathrm{TM})}$ correspond to the transverse electric (TE) and the transverse magnetic (TM) WGMs, respectively, and $\phi$ is the azimuthal angle. Note that the time-dependent electric field as a whole is written as 
\begin{equation}
\bm{\mathsf{E}}(t) = \bm{E}e^{-i \omega t} + \bm{E}^{*} e^{i \omega t}.
\end{equation}

For the clockwise (CW) orbit, the electric fields $\bm{\bar{E}}^{(\mathrm{TE})}$ and $\bm{\bar{E}}^{(\mathrm{TM})}$ can be written as
\begin{eqnarray}
\bm{\bar{E}}^{(\mathrm{TE})} \!\! &=& \left( E^{(m)} e^{im\phi} \right) \Hat{\bm{e}}_{0}^{*},  \label{eq:cTE} \\
\bm{\bar{E}}^{(\mathrm{TM})} \!\! &=& \left( E_{o}^{(m)} e^{im\phi} \right) e^{i \phi} \Hat{\bm{e}}_{+}^{*} \notag \\ &&\hspace{10mm}- \left( E_{i}^{(m)} e^{im\phi} \right) e^{-i \phi} \Hat{\bm{e}}_{-}^{*}. \label{eq:cTM}
\end{eqnarray}

$E_{i}$ ($E_{o}$) in Eqs.~(\ref{eq:TM}) and (\ref{eq:cTM}) shall be called the \textit{inner} (\textit{outer}) component of the TM mode. To see this, Fig.~\ref{Fig:WGM}(a) shows the radial intensity distributions of two components $\left|E_{i} \right|^{2}$ and $\left|E_{o} \right|^{2}$ (magenta and green dotted lines) along with the intensity profiles of the \textit{transverse} component $E_{r}=\frac{1}{\sqrt{2}} \left( E_{i} + E_{o} \right)$ (red solid) and the \textit{longitudinal} component $E_{\mathrm{\phi}}=-\frac{1}{\sqrt{2}} \left( E_{i} - E_{o} \right)$ (blue solid) for the TM electric field of a WGM. We can see that $\left|E_{i} \right|^{2}$ has its maximum in the inner part of the resonator compared to $\left|E_{o} \right|^{2}$. The shift of the ``centers of gravity" of the two components, $\left|E_{i} \right|^{2}$ and $\left|E_{o} \right|^{2}$ is a manifestation of the spin-Hall effect of light~\cite{Murakami, Bliokh1}, which originates from the spin-orbit coupling of light~\cite{Bliokh2}. 

From the dependence of the electric field on $\phi$, the orbital angular momentum $\mathcal{L}_{z}$ of the WGM, that is, the \textit{optical vortex}~\cite{Allen}, can be straightforwardly deduced. First, let us consider the CCW orbit. As for the TE mode with the azimuthal mode index $m=m_{\mathrm{TE}}$, since there is no spin angular momentum, the orbital angular momentum is given by
\begin{equation}
\mathcal{L}_{z}^{(\mathrm{CCW,TE},m_{\mathrm{TE}})} = m_{\mathrm{TE}}. \label{eq:Lz_CCWTE}
\end{equation}
As for the TM mode with $m=m_{\mathrm{TM}}$, however, the spin-orbit coupling of light has to be taken into account~\cite{Bliokh2}.  From the $\phi$-dependence of the coefficient of the first term of Eq.~(\ref{eq:TM}), $\left( E_{i}^{(m_{\mathrm{TM}})} e^{-im_{\mathrm{TM}} \phi} \right) e^{i \phi}$, the orbital angular momentum of the inner ($\sigma_+$) component of the TM mode, $\mathcal{L}_{z}^{(\mathrm{CCW,TM+},m_{\mathrm{TM}})}$, should read as 
\begin{equation}
\mathcal{L}_{z}^{(\mathrm{CCW,TM+},m_{\mathrm{TM}})} = m_{\mathrm{TM}}-1. \label{eq:Lz_CCWTM+}
\end{equation}
From the $\phi$-dependence of the coefficient of the second term of Eq.~(\ref{eq:TM}), $- \left( E_{o}^{(m_{\mathrm{TM}})} e^{-im_{\mathrm{TM}}\phi} \right) e^{-i \phi}$, the orbital angular momentum of the outer ($\sigma_-$) component of the TM mode, $\mathcal{L}_{z}^{(\mathrm{CCW,TM-},m_{\mathrm{TM}})}$, on the other hand, should read as 
\begin{equation}
\mathcal{L}_{z}^{(\mathrm{CCW,TM-},m_{\mathrm{TM}})} = m_{\mathrm{TM}}+1. \label{eq:Lz_CCWTM-}
\end{equation}
Note that since the \textit{spin} angular momentum $\mathcal{S}_{z}^{(+)}=1$ ($\mathcal{S}_{z}^{(-)}=-1$) is associated with the $\sigma_+$ ($\sigma_-$) component [i.e., $\Hat{\bm{e}}_{+}^{*}$ ($\Hat{\bm{e}}_{-}^{*}$) component] of the TM electric field, the \textit{total} angular momentum $\mathcal{J}^{(\mathrm{CCW,TM},m_{\mathrm{TM}})}$ of the TM electric field with azimuthal mode index $m_{\mathrm{TM}}$ is $\mathcal{L}_{z}^{(\mathrm{CCW,TM+},m_{\mathrm{TM}})} + \mathcal{S}_{z}^{(+)}=m_{\mathrm{TM}}$ for the inner  ($\sigma_+$) component and $\mathcal{L}_{z}^{(\mathrm{CCW,TM-},m_{\mathrm{TM}})} + \mathcal{S}_{z}^{(-)}=m_{\mathrm{TM}}$ for the outer ($\sigma_-$) component. Thus, for any cases, $\mathcal{J}^{(\mathrm{CCW,TM},m_{\mathrm{TM}})}=m_{\mathrm{TM}}$ and is well-defined.

For the CW orbit, the similar argument leads us to the following:
\begin{eqnarray}
\mathcal{L}_{z}^{(\mathrm{CW,TE},m_{\mathrm{TE}})} &=& -m_{\mathrm{TE}} \label{eq:Lz_CWTE} \\
\mathcal{L}_{z}^{(\mathrm{CW,TM+},m_{\mathrm{TM}})} &=& -\left( m_{\mathrm{TM}}+1\right) \label{eq:Lz_CWTM+} \\
\mathcal{L}_{z}^{(\mathrm{CW,TM-},M)} &=& -\left( m_{\mathrm{TM}}-1 \right), \label{eq:Lz_CWTM-}
\end{eqnarray}
and the total angular momentum $\mathcal{J}^{(\mathrm{CW,TM},m_{\mathrm{TM}})}=-m_{\mathrm{TM}}$ is again well-defined. Note that for the CW orbit the outer (inner) component of TM mode is associated with $\sigma_+$ ($\sigma_-$), that is opposite to that for the CCW orbit.

The orbital angular momenta of the WGMs can be visualized by sketching the trajectory of the head of the polarization vector of the electric fields [Fig~\ref{Fig:WGM}(b)].  When the mode index is $10$, the orbital angular momentum reads 10, 9 and 11 for the TE, inner TM, and outer TM components, respectively.

\section{Brillouin scattering}

\subsection{Magnetic Quasi-Vortices--Optical Vortices Interaction}
Let us now see that the total orbital angular momentum is conserved in the Brillouin scattering process. The thorough treatment of the Brillouin scattering by magnons in WGMs can be found in Ref.~[16]. In the following, we emphasize the role of orbital angular momenta in the Brillouin scattering process. The interaction Hamiltonian representing the Brillouin scattering is 
\begin{equation}
E = \int \mathcal{E} \mathrm{d}t \mathrm{d}V = \frac{1}{2} \int \bm{E}_2^{*}(t) \tilde{\epsilon}(t) \bm{E}_1(t) \mathrm{d}t \mathrm{d}V, \label{eq:H}
\end{equation}
where the integrand $\mathcal{E}$ is the energy flux density and the integral runs over infinity in time $t$ and the volume $V$ of the WGM, $\bm{E}_1 (t) =\bm{E}_{1} e^{-i \omega_{1} t} $ and $\bm{E}_{2}^{*}(t) =\bm{E}_{2}^{*} e^{i \omega_{2}t} $ are the input and scattered electric fields of WGMs, respectively. Here, the permittivity tensor $\tilde{\epsilon}$ can be written in the Cartesian basis as~\cite{Stancil}
\begin{equation}
\tilde{\epsilon}(t) = \epsilon_0 \left( \epsilon_{r} \bm{\mathsf{M}}_{0} + if M_{x}(t) \bm{\mathsf{M}}_{1} + if M_{y}(t) \bm{\mathsf{M}}_{2} + if M_{s} \bm{\mathsf{M}}_{3} \right) \label{eCart}
\end{equation}
where 
\begin{eqnarray}
\bm{\mathsf{M}}_{0} =\left( \begin{matrix}
		1 & 0 & 0 \\
		0 & 1 & 0 \\
		0 & 0 & 1
		\end{matrix} \right) , \bm{\mathsf{M}}_{1}=\left( \begin{matrix}
		0 & 0 & 0 \\
		0 & 0 & -1 \\
		0 & 1 & 0
		\end{matrix} \right), \notag\\ 
\bm{\mathsf{M}}_{2}=\left( \begin{matrix}
		0 & 0 & 1 \\
		0 & 0 & 0 \\
		-1 & 0 & 0
		\end{matrix} \right), \text{and}\,\, \bm{\mathsf{M}}_{3}=\left( \begin{matrix}
		0 & -1 & 0 \\
		1 & 0 & 0 \\
		0 & 0 & 0
		\end{matrix} \right) \notag
\end{eqnarray}
and $\epsilon_{0}$ is the vacuum permittivity and $\epsilon_{r}$ the relative permittivity. The coefficient $f$ is related to the Verdet constant $\mathcal{V}$ as $f=\frac{2 \sqrt{\epsilon_{r}}}{k_{0} M_{S}} \mathcal{V}$ with the wavevector $k_{0}$ of the optical field in the vacuum~\cite{Stancil}. Here we assumed that the transverse magnetizations $M_{x}$ and $M_{y}$ are time-dependent.

The interaction between the magnetic quasi-vortices and optical vortices in the course of the Brillouin scattering process can be understood best in the spherical basis. In this basis the permittivity tensor can be written as
\begin{equation}
\tilde{\epsilon} = \epsilon_0 \left( \frac{f}{\sqrt{2}} M_{-} e^{i \omega_{m} t} \bm{\mathsf{M}}_{+} +\frac{f}{\sqrt{2}} M_{+}e^{-i \omega_{m} t} \bm{\mathsf{M}}_{-} + f M_{s} \bm{\mathsf{M}}_{z} \right) \label{eSph}
\end{equation}
where $M_{+}e^{-i \omega_{m} t} = M_{x}(t) + i M_{y}(t)$ and $M_{-}e^{i \omega_{m} t} = M_{x}(t) -i M_{y}(t)$, and 
\begin{eqnarray}
&\bm{\mathsf{M}}_{+}=\left( \begin{matrix}
		0 & 1 & 0 \\
		0 & 0 & 1 \\
		0 & 0 & 0
		\end{matrix} \right),
\bm{\mathsf{M}}_{-}=\left( \begin{matrix}
		0 & 0 & 0 \\
		1 & 0 & 0 \\
		0 & 1 & 0
		\end{matrix} \right), \notag\\
&\hspace{10mm} \text{and}\,\, \bm{\mathsf{M}}_{z}=\left( \begin{matrix}
		1 & 0 & 0 \\
		0 & 0 & 0 \\
		0 & 0 & -1
		\end{matrix} \right). \notag
\end{eqnarray}
Here the term $\epsilon_0 \epsilon_r \bm{\mathsf{M}}_{0}$ in Eq.~(\ref{eCart}) has been neglected. Henceforth, the term $f M_s \bm{\mathsf{M}}_{z}$ in Eq.~(\ref{eSph}) is also ignored for it is independent of time and give no contribution to the Brillouin scattering. Here $\omega_{m}/2\pi$ is the resonant frequency of the concerned Walker mode with the azimuthal mode index of $m_{\mathrm{mag}}$.

In the spherical basis the time-dependent transverse magnetization is given by
\begin{equation}
\bm{\mathsf{M}}(t) = M_{-}e^{i \omega_{m} t} \bm{\mathsf{M}}_{+} + M_{+}e^{-i \omega_{m} t} \bm{\mathsf{M}}_{-}.
\end{equation}
Here note that the creation (annihilation) of a magnon \textit{decreases} (\textit{increases}) the spin angular momentum. As we shall show, the Brillouin scattering stems from the term with $\bm{\mathsf{M}}_{+}$ ($\bm{\mathsf{M}}_{-}$) representing the Stokes-scattering (anti-Stokes-scattering) associated with the creation (annihilation) of a magnon. Since the TE-to-TM or TM-to-TE transition process changes the spin angular momentum of magnon, these transitions give nonzero contributions to the Brillouin scattering given the conservation of the spin angular momentum. We shall see this more clearly in Sec.~\ref{sec:sr}.

\subsection{Selection rules} \label{sec:sr}

Since the interaction depends on the direction of the input field and its polarization, let us first suppose that the input field is the CW TE mode with mode index of $m_{\mathrm{TE}}$, that is, $\bm{E}_{1}(t) = E^{(m_{\mathrm{TE}}) }e^{-im_{\mathrm{TE}}\phi}e^{-i \omega_{1}t} \Hat{\bm{e}}_{0}^{*}$. In this case the Brillouin scattering results in producing photons in the CW TM mode as seen in the following. We can straightforwardly extend the argument to other cases, e.g., the TM mode input or the input to the CCW orbit. 

With the CW TE mode as the input field, the energy flux density $\mathcal{E}$ in Eq.~(\ref{eq:H}) reads
\begin{eqnarray}
&\ &\!\!\! \mathcal{E} = \frac{\epsilon_0 f}{2 \sqrt{2}} \sum_{m_{\mathrm{TM}}}  \left(M_{-} E_{o}^{(m_{\mathrm{TM}}) *} E^{(m_{\mathrm{TE}})} e^{i \Delta\mathcal{L}_{+} \phi} e^{i \delta_{+} t} \right. \nonumber \\
&\ & \hspace{15mm} \left. \ \ - M_{+} E_{i}^{(m_{\mathrm{TM}}) *} E^{(m_{\mathrm{TE}})} e^{i \Delta\mathcal{L}_{-} \phi} e^{i \delta_{-} t} \right), \label{eq:Ed}
\end{eqnarray}
where
\begin{eqnarray}
\Delta\mathcal{L}_{\pm} &=& \mathcal{L}_{z}^{(\mathrm{CW,TM\pm},m_{\mathrm{TM}})}\notag \\ &&\hspace{5mm} - \mathcal{L}_{z}^{(\mathrm{CW,TE},m_{\mathrm{TE}})} \pm \mathcal{L}_{z}^{(m_{\mathrm{mag}})}, \\
\delta_{\pm} &=& \omega_2 - \omega_1 \pm \omega_m.
\end{eqnarray}
The first (second) term in the right-hand side of Eq.~(\ref{eq:Ed}) represents the Stokes (anti-Stokes) scattering. The possibility of the scattered light being the CCW WGM is denied given the fact that we are concerned only with cases where $\mathcal{L}_{z}^{(m_{\mathrm{mag}})} \ll \mathcal{L}_{z}^{(\mathrm{TE})}, \mathcal{L}_{z}^{(\mathrm{TM})}$.

The integration with respect to time $t$ in Eq.~(\ref{eq:H}) with Eq.~(\ref{eq:Ed}) leads to the energy conservation 
\begin{equation}
\omega_{2} - \omega_{1} + \omega_{m} =0 \label{eq:S}
\end{equation}
for the Stokes scattering and
\begin{equation}
\omega_{2} - \omega_{1} - \omega_{m}=0 \label{eq:AS}
\end{equation}
for the anti-Stokes scattering. Since the optical densities of states are modified in the presence of the WGMs, the probabilities of the scattering processes are affected by them, too. 

Furthermore, because of the axial symmetry of the system, the conservation of the total angular momentum is expected. The designated WGM of the Brillouin scattering can then be specified by the selection rule obtained by the conservation of the orbital angular momentum. To see this, we integrate $\mathcal{E}$ in Eq.~(\ref{eq:H}) over the azimuthal angle $\phi$ as a part of the volume integral. From the first Stokes term in Eq.~(\ref{eq:Ed}) we have a selection rule:
\begin{equation}
 \mathcal{L}_{z}^{(\mathrm{CW,TM+},m_{\mathrm{TM}})} - \mathcal{L}_{z}^{(\mathrm{CW,TE},m_{\mathrm{TE}})} + \mathcal{L}_{z}^{(m_{\mathrm{mag}})} = 0.
\end{equation}
With Eqs.~(\ref{eq:Lz_m}), (\ref{eq:Lz_CWTE}), and (\ref{eq:Lz_CWTM+}), this selection rule amounts to
\begin{equation}
m_{\mathrm{TM}} = m_{\mathrm{TE}} - m_{\mathrm{mag}}. \label{eq:CW_TE_TM+}
\end{equation}
As for the second anti-Stokes term in Eq.~(\ref{eq:Ed}), the selection rule is 
\begin{equation}
 \mathcal{L}_{z}^{(\mathrm{CW,TM-},m_{\mathrm{TM}})} - \mathcal{L}_{z}^{(\mathrm{CW,TE},m_{\mathrm{TE}})} - \mathcal{L}_{z}^{(m_{\mathrm{mag}})} = 0,
\end{equation}
and with Eqs.~(\ref{eq:Lz_m}), (\ref{eq:Lz_CWTE}), and (\ref{eq:Lz_CWTM-}), we have
\begin{equation}
m_{\mathrm{TM}} = m_{\mathrm{TE}} + m_{\mathrm{mag}}. \label{eq:CW_TE_TM-}
\end{equation}

Next, let us briefly describe the results when the laser light is injected into the CCW-TE mode. The Stokes (anti-Stokes) scattering process gives the same conditions of the energy conservation, Eq.~(\ref{eq:S}) [Eq.~(\ref{eq:AS})].  However, for the CCW case the Stokes (anti-Stokes) scattering originates in the inner (outer) component in contrast to the CW case.  For the anti-Stokes scattering, the conservation of the orbital angular momentum is represented by 
\begin{equation}
\mathcal{L}_{z}^{(\mathrm{CCW,TM+},m_{\mathrm{TM}})} - \mathcal{L}_{z}^{(\mathrm{CCW,TE},m_{\mathrm{TE}})} - \mathcal{L}_{z}^{(m_{\mathrm{mag}})} = 0.
\end{equation}
which results in
\begin{equation}
m_{\mathrm{TM}} = m_{\mathrm{TE}} - m_{\mathrm{mag}}, \label{eq:CCW_TE_TM+}
\end{equation}
that is, the same selection rule as Eq.~(\ref{eq:CW_TE_TM+}). As for the Stokes scattering, 
\begin{equation}
\mathcal{L}_{z}^{(\mathrm{CCW,TM-},m_{\mathrm{TM}})} - \mathcal{L}_{z}^{(\mathrm{CCW,TE},m_{\mathrm{TE}})} + \mathcal{L}_{z}^{(m_{\mathrm{mag}})} = 0,
\end{equation}
represents the orbital angular momentum conservation, yielding 
\begin{equation}
m_{\mathrm{TM}} = m_{\mathrm{TE}} + m_{\mathrm{mag}}, \label{eq:CCW_TE_TM-}
\end{equation}
that is, the same selection rule as Eq.~(\ref{eq:CW_TE_TM-}).  

{These selection rules regarding the orbital angular momentum are the main result of this paper. With the geometric birefringence~\cite{SB, Lam, Schiller} and densities of states of WGMs, these selection rules dictate the Brillouin light scattering by Walker-mode magnons hosted in a ferromagnetic sphere as shown below. In the next section we employ the selection rules to explain the experiment reported in the accompanying paper~\cite{PRL}, which reveals that the Walker-mode-induced Brillouin light scattering is either nonreciprocal or reciprocal depending on the orbital angular momentum of the magnon in the relevant Walker mode, that is, the magnetic quasi-vortex.

\begin{figure}[t]
\includegraphics[width=8.5cm]{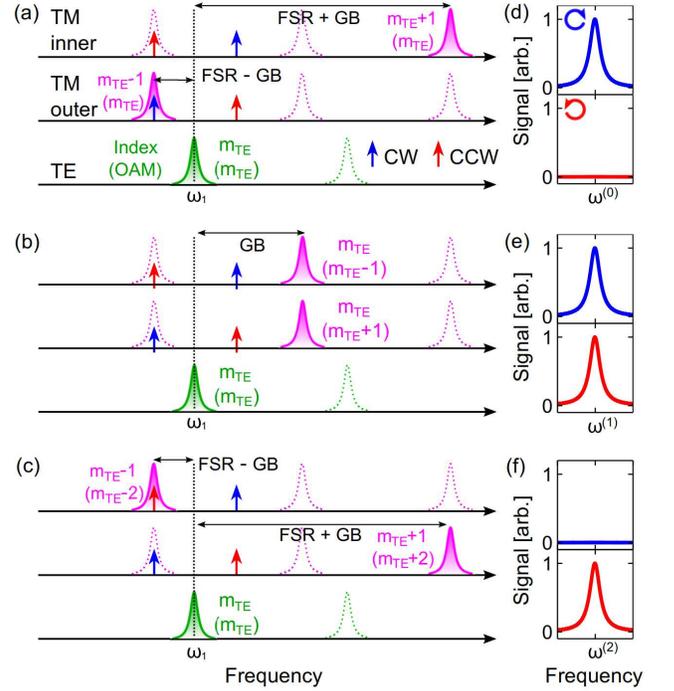}
\caption{\label{FigAnalysis} (a)-(c): Relevant TE (green) and TM (purple) WGMs in the Brillouin scattering (see main text) with the orbital angular momenta (OAM) of the Walker modes being (a) $0$, (b) $1$ and (c) $2$. The mode indices and the orbital angular momentum of the relevant WGMs are indicated by the labels next to the Lorentzian peaks.  FSR and GB stand for the free spectral range and the frequency shift due to the geometric birefringence, respectively.  (d)-(f): Theoretically predicted reciprocal/nonreciprocal behavior of the Brillouin scattering signals corresponding to the cases (a)-(c), respectively.  The frequency $\omega^{(i)}$ represents that of the Walker mode with orbital angular momentum of $i$.}
\end{figure}

\section{Nonreciprocal Brillouin scattering}

Here, we apply the selection rules to a concrete example. We suppose that the sample under consideration is a sphere made of yttrium iron garnet (YIG) with its diameter of $1$~mm, and focus on the Brillouin scattering by the Walker-mode magnons with $\mathcal{L}_z^{(m_{\mathrm{mag}})}=0, 1, 2$, which include $(1, 1, 0)$, $(3, 1, 1)$, $(4, 0, 1)$ and $(3, \bar{1}, 1)$ Walker modes experimentally investigated in the accompanying paper~\cite{PRL}.

\subsection{Walker modes with $\mathcal{L}_z^{(m_{\mathrm{mag}})} = 0$}

We first consider the simplest family of the Walker modes, namely, those having $\mathcal{L}_z^{(m_{\mathrm{mag}})} = 0$, that is, $m_{\mathrm{mag}} = 1$.  Equation~(\ref{eq:CW_TE_TM+}) reduces to $m_{\mathrm{TM}} = m_{\mathrm{TE}} - 1$ and Eq.~(\ref{eq:CW_TE_TM-}) to $m_{\mathrm{TM}} = m_{\mathrm{TE}} + 1$, corresponding respectively to the cases of the Stokes and the anti-Stokes scattering with the CW-TE mode input. On the other hand, Eq~(\ref{eq:CCW_TE_TM+}) reduces to $m_{\mathrm{TM}} = m_{\mathrm{TE}} - 1$ and Eq.~(\ref{eq:CCW_TE_TM-}) to $m_{\mathrm{TM}} = m_{\mathrm{TE}} + 1$, corresponding respectively to the cases of the anti-Stokes and the Stokes scattering with the CCW-TE mode input. The schematics of the relevant scattering processes are depicted in the Fig.~\ref{FigAnalysis}(a).

In the figure, the densities of states of the TE (green) and the TM (purple) WGMs are schematically shown.  Note that the TE and the TM WGM resonances does not have the same frequency due to the geometric birefringence~\cite{SB, Lam, Schiller}.  The difference between these two frequencies is denoted by GB in the figure, which is about 0.9 times the free spectral range ($\mathrm{FSR}$) in a spherical resonator made of YIG.  Two sets of the TM modes are depicted (top and middle), one for the inner and the other for the outer components.   Hereafter, for clarity of the analysis, the light is supposed to be injected into the TE WGM with the mode index $m_{\mathrm{TE}}$ (green highlighted).   

The relevant TM WGMs specified by the selection rules are highlighted (otherwise dotted) in each of the inner or the outer case in Fig.~\ref{FigAnalysis}(a).  The frequencies that the scattered light would acquire are indicated by blue and red upright arrows for the CW and the CCW cases, respectively.  Here we made an assumption that the difference of the input and scattered light determined by the Walker-mode frequency is tuned to coincide with the value $\mathrm{FSR}-\mathrm{GB}$ by the applied magnetic field, which is experimentally feasible as it is approximately realized in the experiment in the accompanying paper~\cite{PRL} as well as the others~\cite{Tang2016,Haigh2016}. For the case of the scattering into the inner component of TM mode (top panel), the scattered light for both the CW and the CCW cases are far detuned from the selection-rule-allowed WGM (highlighted). On the other hand, for the case of the scattering into the outer component of TM mode (middle panel) the scattered light is almost resonant to the selection-rule-allowed WGM (highlighted) for the CW case but off-resonant for the CCW case. Hence we can conclude that for Walker modes with $\mathcal{L}_z^{(m_{\mathrm{mag}})} = 0$ ($m_{\mathrm{mag}}=1$) the Brillouin scattering of the CW WGM is expected to be more intense than that of the CCW case. The expected nonreciprocal behavior of the Brillouin scattering is schematically shown in Fig.~\ref{FigAnalysis}(d), which is in agreement with the experimentally observed Brillouin scattering signals by the $(1, 1, 0)$ and the $(3, 1, 1)$ Walker modes~\cite{PRL}.

\subsection{Walker modes with $\mathcal{L}_z^{(m_{\mathrm{mag}})} = 1$}

Next we consider the Walker mode with $\mathcal{L}_z^{(m_{\mathrm{mag}})} = 1$ ($m_{\mathrm{mag}} = 0$). In this case all the selection rules, Eqs.~(\ref{eq:CW_TE_TM+}), (\ref{eq:CW_TE_TM-}),(\ref{eq:CCW_TE_TM+}), and (\ref{eq:CCW_TE_TM-}), amount to the same condition $m_{\mathrm{TM}} = m_{\mathrm{TE}}$.  In other words, the selection rules are the same for both the CCW and CW orbits, resulting in the absence of the nonreciprocity [Fig.~\ref{FigAnalysis}(e)].  In Ref.~[22], the $(4, 0, 1)$ mode corresponding to this case actually exhibits the reciprocal behavior.

\subsection{Walker modes with $\mathcal{L}_z^{(m_{\mathrm{mag}})} = 2$}

As the final example, let us examine the Walker mode with $\mathcal{L}_z^{(m_{\mathrm{mag}})} = 2$ ($m_{\mathrm{mag}} = -1$). The selection rules are obtained by inserting $m_{\mathrm{mag}} = -1$ into Eqs.~(\ref{eq:CW_TE_TM+}), (\ref{eq:CW_TE_TM-}), (\ref{eq:CCW_TE_TM+}), and (\ref{eq:CCW_TE_TM-}).  Equation~(\ref{eq:CW_TE_TM+}) yields $m_{\mathrm{TM}} = m_{\mathrm{TE}} + 1$ for the Stokes scattering of the CW WGM, and Eq.~(\ref{eq:CW_TE_TM-}) yields $m_{\mathrm{TM}} = m_{\mathrm{TE}} - 1$ for the anti-Stokes scattering of the CW WGM, while Equation~(\ref{eq:CCW_TE_TM+}) yields $m_{\mathrm{TM}} = m_{\mathrm{TE}} + 1$ for the anti-Stokes scattering of the CCW WGM, and Eq.~(\ref{eq:CCW_TE_TM-}) yields $m_{\mathrm{TM}} = m_{\mathrm{TE}} - 1$ for the Stokes scattering of the CCW WGM. Here, the situations shown in Fig.~\ref{FigAnalysis}(c) are opposite to that shown in Fig.~\ref{FigAnalysis}(a) for $\mathcal{L}_z^{(m_{\mathrm{mag}})} = 0$. Consequently, the Brillouin scattering of the CCW WGM would show larger signal than that of the CW WGM as indicated in Fig.~\ref{FigAnalysis}(f), again agreeing well with the experimental result~\cite{PRL}.

\section{Conclusion} \label{5}

We analyzed the spin and orbital angular momenta of magnons in the Walker modes and photons in the whispering gallery modes, both of which being supported by a ferromagnetic sphere. We then predicted that in the Brillouin light scattering within the ferromagnetic sphere the orbital angular momenta are exchanged between the photons and the magnons in such a way that the total orbital angular momentum is conserved. The nontrivial nonreciprocal/reciprocal behaviors in the Brillouin scattering by the Walker-mode magnons observed in the experiment reported in the accompanying paper~\cite{PRL} were then explained as the consequence of the selection rule imposed by the conservation of the total orbital angular momentum. Our findings will stimulate further investigation on the role of the orbital angular momenta in cavity optomagnonics, e.g., in view of chiral quantum optics~\cite{Lodahl2017} and topological photonics~\cite{Taylor2014,Soljacic2014}.

We are grateful to R.~Hisatomi, A.~Noguchi, R.~Yamazaki, M.~Nomura, Y.~Shikano, Y.~Tabuchi, Y.~Nakata, S.~Ishino, G.~Tatara, S.~Iwamoto, J.~Haigh, S. Sharma, Y.~M.~Blanter, and  G.~E.~W.~Bauer for fruitful discussions. This work was supported by JSPS KAKENHI (Grant No.~15H05461, No.~16F16364, No.~26220601, and No.~26600071), Murata Science Foundation, the Inamori Foundation, and JST ERATO project (Grant No. JPMJER1601).  AG is an Overseas researcher under Postdoctoral Fellowship of JSPS.



\end{document}